\documentclass[10pt,twocolumn,letterpaper]{article}

\usepackage[pagenumbers]{iccv} % To force page numbers, e.g. for an arXiv version
\usepackage[letterpaper,top=2cm,bottom=2cm,left=3cm,right=3cm,marginparwidth=1.75cm]{geometry}
\usepackage{multirow}
\definecolor{iccvblue}{rgb}{0.21,0.49,0.74}
\usepackage[pagebackref,breaklinks,colorlinks,allcolors=iccvblue]{hyperref}
\usepackage{amsmath}
\usepackage{amssymb}
\usepackage{graphicx}
 \usepackage{booktabs}
\usepackage{float}
\usepackage{subcaption}
\usepackage[font=small,skip=0pt]{caption}
\def\paperID{5} % *** Enter the Paper ID here
\def\confName{ICCV}
\def\confYear{2025}

\title{DinoAtten3D: Slice-Level Attention Aggregation of DinoV2 for 3D Brain MRI Anomaly Classification}

\author{
Fazle Rafsani\\
Arizona State University\\
{\tt\small frafsani@asu.edu}
\and
Jay Shah\\
Arizona State University\\
{\tt\small jgshah1@asu.edu}
\and
Catherine D. Chong\\
Mayo Clinic, Arizona\\
{\tt\small Chong.Catherine@mayo.edu}
\and
Todd J. Schwedt\\
Mayo Clinic, Arizona\\
{\tt\small schwedt.todd@mayo.edu}
\and 
Teresa Wu\\
Arizona State University\\
{\tt\small teresa.wu@asu.edu}
}

\begin{document}
\maketitle
\begin{abstract}
Anomaly detection and classification in medical imaging are critical for early diagnosis but remain challenging due to limited annotated data, class imbalance, and the high cost of expert labeling. Emerging vision foundation models such as DINOv2, pretrained on extensive, unlabeled datasets, offer generalized representations that can potentially alleviate these limitations. In this study, we propose an attention-based global aggregation framework tailored specifically for 3D medical image anomaly classification. Leveraging the self-supervised DINOv2 model as a pretrained feature extractor, our method processes individual 2D axial slices of brain MRIs, assigning adaptive slice-level importance weights through a soft attention mechanism. To further address data scarcity, we employ a composite loss function combining supervised contrastive learning with class-variance regularization, enhancing inter-class separability and intra-class consistency. We validate our framework on the ADNI dataset and an institutional multi-class headache cohort, demonstrating strong anomaly classification performance despite limited data availability and significant class imbalance. Our results highlight the efficacy of utilizing pretrained 2D foundation models combined with attention-based slice aggregation for robust volumetric anomaly detection in medical imaging. Our implementation is publicly available at \hyperlink{https://github.com/Rafsani/DinoAtten3D.git}{https://github.com/Rafsani/DinoAtten3D.git}.
\end{abstract}

\section{Introduction}

Anomaly detection and classification in medical imaging pose significant challenges due to data scarcity and the high cost associated with obtaining expert annotations. Traditionally, supervised methods have been widely adopted, but these approaches demand extensive labeled datasets and are susceptible to overfitting, especially in scenarios characterized by class imbalance and limited sample sizes. Consequently, unsupervised methods have emerged as alternative solutions, employing reconstruction-based frameworks utilizing generative adversarial networks (e.g., HealthyGAN \cite{rahman2022healthygan}, Brainomaly \cite{siddiquee2024brainomaly}, f-AnoGAN \cite{schlegl2017fast}) or diffusion models (e.g., AnoDDPM \cite{wyatt2022anoddpm}, AnoFPDM \cite{che2025anofpdm}). These unsupervised methods reconstruct healthy versions of potentially anomalous images and detect anomalies through deviations, yet they may also suffer from overfitting due to limited healthy training data, hindering generalization and performance \cite{varoquaux2022machine, zhang2022unsupervised}.

The recent rise of large-scale foundation models offers a potential remedy to these fundamental limitations. Models such as GPT \cite{radford2018improving}, CLIP \cite{radford2021learning}, and DINOv2 \cite{oquab2023dinov2} achieve remarkable performance by extensively pretraining on large and heterogeneous datasets, thus effectively internalizing broad statistical regularities across language and vision modalities \cite{wang2025scaling}. These foundation models exhibit emergent in-context learning abilities, enabling effective zero- or few-shot learning by leveraging generalized latent representations \cite{bommasani2021opportunities, liu2024few}.

In medical imaging, multimodal foundation models rely on paired image-text data. However, producing high-quality clinical captions is often impractical and laborious. This has motivated the development of vision-only models pretrained exclusively on medical imaging data. DINOv2 exemplifies this paradigm, demonstrating robust generalization as a feature extractor across diverse vision tasks, including medical imaging data \cite{baharoon2023evaluating}.

Despite these advantages, foundation models like DINOv2 are inherently designed for 2D image processing and cannot natively handle volumetric medical images such as 3D MRI scans that are ubiquitous in clinical practice. Slice-based methods have therefore been proposed, independently processing 2D slices from volumetric data. Additionally, multi-instance learning (MIL) approaches, traditionally applied to high-dimensional data like whole-slide images (WSIs), segment large images into smaller instances or patches \cite{campanella2019clinical, ilse2018attention, yan2023mil3d}. While effective in identifying local discriminative regions, traditional MIL methods may overlook broader spatially distributed pathological features across multiple slices that are essential for accurate diagnosis \cite{lu2021data, yan2023mil3d}.

To address these limitations, we propose a DINOv2-based soft attention-driven global aggregation approach tailored specifically for 3D medical imaging (\textbf{DinoAtten3D}) that overcomes the inherent dimensionality constraints of 2D foundation models. It leverages DINOv2's rich embeddings from axial slices of 3D MRI volumes and introduces a soft attention mechanism that adaptively weighs whole-slice (instead of patch) embeddings based on their diagnostic relevance. By emphasizing diagnostically significant slices, our approach efficiently captures both focal and distributed pathological patterns, bridging the gap between the inherent 2D capabilities of DINOv2 and the volumetric nature of medical imaging, without relying on computationally intensive 3D models that are often impractical for clinical deployment. In summary, our main contributions include:
% \begin{itemize}
% \item A global attention-based aggregation framework for 3D medical image analysis, adaptively fusing slice-level embeddings using soft attention pooling to highlight diagnostically informative slices.
% \item Validation of our method on two clinical datasets: the Alzheimer's Disease Neuroimaging Initiative (ADNI) and a multi-class headache cohort, demonstrating robust performance despite data scarcity and class imbalance.
% \item Effective classification not only between healthy and pathological cases but also among various pathological subtypes, demonstrating broader applicability in clinical diagnostics.
% \end{itemize}

\begin{itemize}
    \item We propose a global attention-based aggregation framework for 3D medical imaging that adaptively fuses slice-level embeddings from all 2D slices via soft attention pooling. This highlights the most informative slices while retaining distributed pathological cues, enabling volumetric classification using lightweight 2D backbones such as DINOv2, without the computational overhead of fully 3D architectures.
    \item We validate our approach on two real-world clinical cohorts: Alzheimer’s Disease Neuroimaging Initiative (ADNI) MRI and a multi-class headache cohort, showing strong anomaly classification performance even under severe data scarcity and class imbalance.
    \item Beyond differentiating unhealthy from healthy samples, our framework also achieves promising results in distinguishing between different pathological subtypes, demonstrating its broader utility for downstream clinical analyses.
\end{itemize}

\section{Related Works}

Anomaly detection in medical imaging has garnered significant research attention, leading to a wide range of methodologies for identifying pathological deviations~\cite{tschuchnig2021anomaly}.  
Early work focused primarily on supervised learning approaches, particularly Convolutional Neural Networks (CNNs), which are trained using annotated datasets to learn discriminative representations of anomalies~\cite{paul2024efficient, mortazi2018automatically, li2014medical, salehi2023study, deepak2019brain, kayalibay2017cnn, iqbal2018brain, havaei2017brain, shin2016deep}. While CNN-based models have achieved success in both classification and segmentation tasks, their reliance on large, expert-labeled datasets poses significant challenges given the high cost and labor intensity of medical image annotation.

To alleviate the annotation bottleneck, unsupervised and self-supervised approaches have emerged. For example, Reconstruction-based anomaly detection methods leverage healthy images to learn normative distributions, flagging deviations at inference as potential anomalies. Generative adversarial networks (GANs), such as f-AnoGAN~\cite{schlegl2017fast}, Brainomaly~\cite{siddiquee2024brainomaly}, and HealthyGAN~\cite{rahman2022healthygan}, generate healthy counterfactuals of test images to detect anomalies. More recently, diffusion-based methods~\cite{dhariwal2021diffusion, wyatt2022anoddpm, che2025anofpdm, mykula2024diffusion} have gained traction due to superior generative performance; however, these models still require substantial amounts of healthy training data, and efforts are still being made to address this issue with these models~\cite{wang2023patch, daras2024how, li2024pruning}.

Medical imaging data are often high-dimensional and can possess complex three-dimensional structures, as seen in modalities such as MRI and CT. Whole Slide Images (WSIs), commonly used in cancer diagnosis, are particularly large and high-resolution. To efficiently process these data, patch-based training strategies and instance-based learning methods are frequently employed~\cite{ciga2021overcoming, lu2021data}.  
In this context, multi-instance learning (MIL) has become a popular paradigm, especially in digital pathology and large-scale medical image analysis~\cite{campanella2019clinical, ilse2018attention, lu2021data, yan2023mil3d}. Traditional MIL approaches divide images into patches or instances and use aggregation strategies such as max-pooling, mean-pooling, or attention pooling to generate bag-level predictions from instance-level features \cite{carbonneau2018multiple}. Attention-based MIL~\cite{ilse2018attention} is particularly effective, as it enables the model to learn which regions are most informative for downstream tasks. These methods, however, often focus on local or spatially sparse regions (e.g., tumor), which may not capture global or distributed pathological patterns, especially in volumetric imaging where relevant features may be distributed across multiple slices~\cite{crespi20223d}. And, later, is particularly true to some neurodegenerative diseases such as Alzheimer's and headaches.

To address class imbalance in anomaly detection and enhance representation learning in MIL frameworks, recent works such as SC-MIL (Supervised Contrastive Multiple Instance Learning)~\cite{li2021scmil} have introduced supervised contrastive loss to promote more discriminative and robust bag-level embeddings. SC-MIL is particularly relevant as it addresses the challenge of imbalanced classification, a common issue in medical imaging, and forms a strong baseline for our work due to its close methodological proximity to our approach, differing mainly in its use of local instance aggregation compared to our proposed global attention-based aggregation strategy.

Amidst these developments, foundation models have demonstrated remarkable generalizability across diverse vision and language tasks by leveraging large-scale pretraining~\cite{zhang2024challenges}. However, developing foundation models specifically for medical imaging remains challenging due to data scarcity and the cost of collecting paired image-text data. Vision-language foundation models such as BioMedCLIP~\cite{zhang2023biomedclip} and MedSAM~\cite{ma2024segment} leverage contrastive learning and prompts. Researchers are increasingly exploring vision-only models pretrained on natural images for medical imaging tasks~\cite{huix2024natural}. Recent work has shown that DINOv2, a self-supervised vision transformer pretrained on large natural image datasets, provides robust and transferable feature representations, outperforming other pre-trained models in medical image classification~\cite{oquab2023dinov2, baharoon2023evaluating}. While DINOv2 is inherently 2D, its strong feature extraction capabilities make it an attractive choice for 3D medical imaging when paired with effective aggregation strategies.

\par
In summary, our approach is motivated by (1) the proven effectiveness of attention pooling for large-scale and high-dimensional medical image analysis, (2) recent advances in supervised contrastive learning for imbalanced classification as exemplified by SC-MIL, and (3) the demonstrated generalizability of foundation models such as DINOv2 as feature extractors. Our approach contrasts with conventional local/instance-level MIL by proposing a global, attention-driven aggregation of slice-level DINOv2 embeddings that is specifically designed to address the different challenges of 3D medical images like brain MRI.

\section{Method}
In this section, we present the training procedure of the model and the architecture of the attention pooling-based 3D brain MRI classification task.
\subsection{The Model Architecture: DinoAtten3D}
\begin{figure*}
    \centering
    \includegraphics[width=0.9\linewidth]{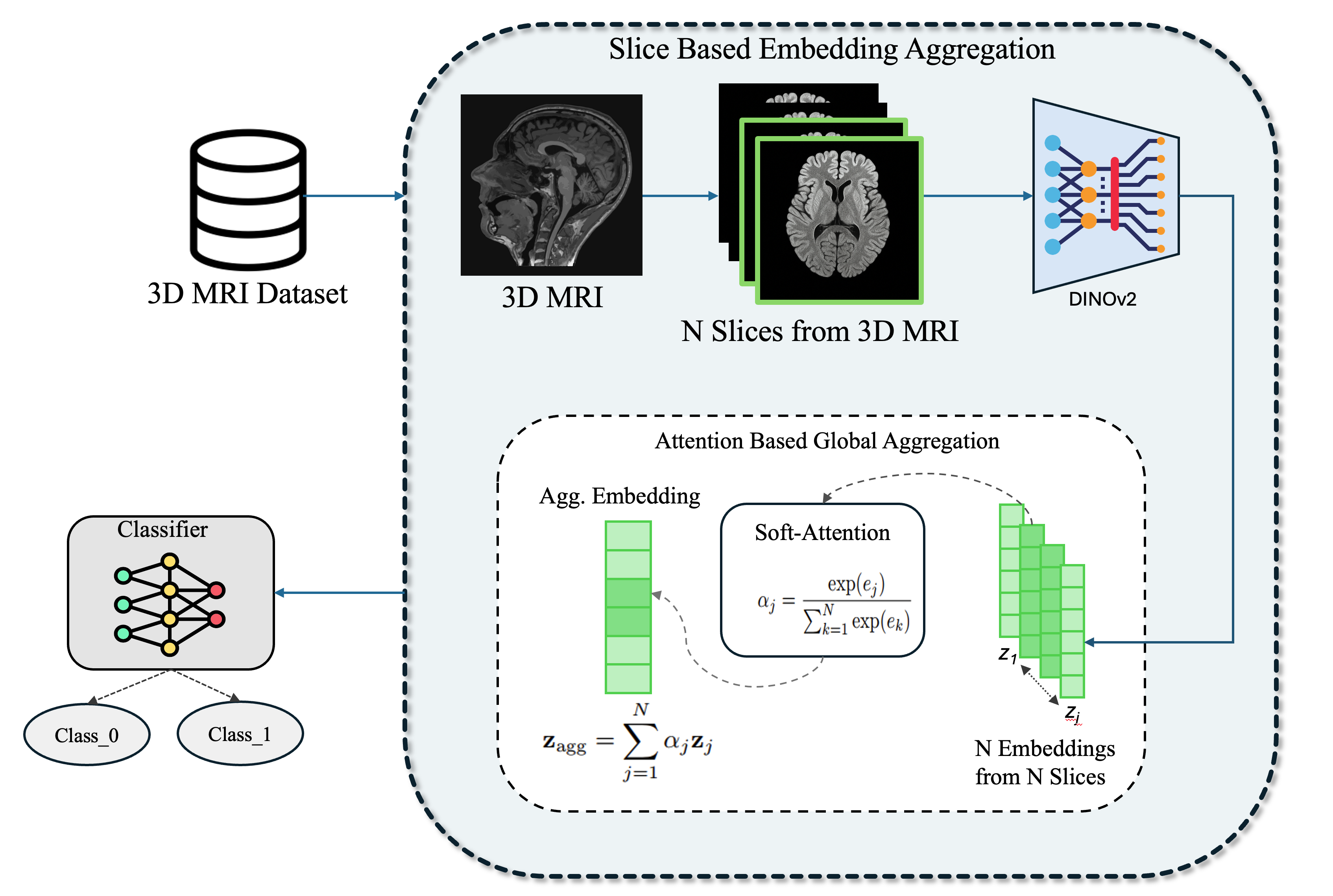}
    \caption{Overall architecture of the slice-based attention aggregation of 2D slice embeddings for 3D brain MRI}
    \label{fig:main_method}
\end{figure*}

The DinoAtten3D comprises a pre-trained feature extractor, an attention-pooled weighted aggregation block, and an MLP (Multi-Layer Perceptron) on top of the attention block. The full overview of the method is presented in Figure \ref{fig:main_method}. The foundation model, DinoV2, trained through a self-supervised process, is used on 2D slices of the MRI volume to extract rich embeddings for each slice. Later, these 2D slice embeddings are given an attention score, further explained in section \ref{method:slice-agg}. The cumulative weighted embeddings are then used to train a classifier. The training process employs a custom loss function that combines cross-entropy loss, contrastive loss, and a class variance loss (see section~\ref{within-class}) to effectively learn from a relatively small dataset. While cross-entropy loss guides the model to correctly classify samples based on ground truth labels, it alone may not be sufficient to capture subtle intra-class variations or enhance feature separability, particularly in low-data regimes. To address this, contrastive loss is integrated to encourage the model to learn discriminative representations by pulling together embeddings of samples from the same class and pushing apart those from different classes in the feature space. Additionally, the class variance loss is introduced to minimize the intra-class variance, ensuring that embeddings of samples within the same class remain compact. 

\subsubsection{2D Feature Extraction via Frozen DinoV2}

Let a 3D MRI volume be represented by an ordered set of $N$ axial slices, where $\mathcal{S}$ represents the set of slices for each 3D MRI volume.
\[
  \mathcal{S} = \{\,S_j \in \mathbb{R}^{C \times H \times W} \mid j = 1, \dots, N\}.
\]
Each slice $S_j$ is passed through the pretrained, frozen DinoV2 backbone 
\[
f_{\mathrm{Dino}}: \mathbb{R}^{C \times H \times W} \rightarrow \mathbb{R}^d,
\]
where $d=384$, producing per‐slice embeddings. In our work, we utilize the ViT‑based backbone of DINOv2 (often denoted ViT‑S/14), which tokenizes each input slice into a sequence of \(14 \times 14\) patches and projects them into a 384‑dimensional latent space. So
\[
  \mathbf{z}_j = f_{\mathrm{Dino}}(S_j)
  \in \mathbb{R}^d,
  \quad j = 1, \dots, N.
\]

\subsubsection{Slice‐Level Global Attention Aggregation}
\label{method:slice-agg}
To aggregate the $N$ slice embeddings into a single volume-level feature, we learn scalar attention scores with a two-layer MLP:
\[
e_j = \mathbf{w}_2^\top \tanh(\mathbf{W}_1 \mathbf{z}_j)
\]
with $\mathbf{W}_1 \in \mathbb{R}^{h \times d}$, $\mathbf{w}_2 \in \mathbb{R}^h$. Attention weights are then obtained via softmax:
\[
\alpha_j = \frac{\exp(e_j)}{\sum_{k=1}^N \exp(e_k)}
\]
and the aggregated feature is:
\[
\mathbf{z}_{\mathrm{agg}} = \sum_{j=1}^N \alpha_j \mathbf{z}_j.
\]

\subsubsection{Embedding Head and Classification}
The aggregated embedding $\mathbf{z}_{\mathrm{agg}}$ is mapped to a lower‐dimensional embedding via a two‐layer MLP.
% \[
% h(\mathbf{z}) = \mathrm{ReLU}\!\bigl(\mathbf{W}_4\,\mathrm{ReLU}(\mathbf{W}_3\,\mathbf{z})\bigr),
% \]
% with $\mathbf{W}_3 \in \mathbb{R}^{p \times d}$ and $\mathbf{W}_4 \in \mathbb{R}^{m \times p}$, where $p=256$ and $m=128$.  Denote
% \begin{equation}
%   \mathbf{h} = h(\mathbf{z}_{\mathrm{agg}}) \in \mathbb{R}^m.
% \end{equation}
Finally, a linear classifier produces logits for the binary classification.
% \begin{equation}
%   \mathbf{o} = c(\mathbf{h}) = \mathbf{W}_5\,\mathbf{h} + \mathbf{b}_5,
%   \quad \mathbf{o} \in \mathbb{R}^2.
% \end{equation}

\subsection{Training Objective}
Let $\tilde{\mathbf{h}}^{(i)} = \mathbf{h}^{(i)} / \|\mathbf{h}^{(i)}\|_2$ be the normalized embedding for sample~$i$ in a batch of size $B$, and let $y^{(i)}$ be its class label. The overall loss combines three terms: a standard cross-entropy loss on classifier logits, a supervised contrastive loss on embeddings, and a within-class variance regularization.

\medskip
\noindent\textbf{Cross-Entropy Loss:}
Given predicted logits $\mathbf{o}^{(i)} \in \mathbb{R}^C$, the cross-entropy loss is:
\[
  \mathcal{L}_\mathrm{CE}
  = -\frac{1}{B} \sum_{i=1}^B \log \bigl( \mathrm{softmax}(\mathbf{o}^{(i)})_{y^{(i)}} \bigr).
\]

\medskip
\noindent\textbf{Contrastive Loss:}
We compute pairwise similarities between normalized embeddings:
\[
  s_{ij} = \frac{1}{\tau} \tilde{\mathbf{h}}^{(i)\,\mathsf{T}} \tilde{\mathbf{h}}^{(j)},
\]
where $\tau > 0$ is a temperature parameter. Let $\mathcal{P}(i)$ be the set of indices sharing the same label as sample~$i$. The contrastive loss is:
\[
  \mathcal{L}_\mathrm{contra}
  = \frac{1}{B} \sum_{i=1}^B \left[
      -\frac{1}{|\mathcal{P}(i)|} \sum_{j \in \mathcal{P}(i)} 
      \log \frac{\exp(s_{ij})}{\sum_{k \neq i} \exp(s_{ik})}
    \right].
\]

\medskip
\noindent\textbf{Within-Class Variance Loss:}
\label{within-class}
To encourage compact clustering of same-class embeddings, we define for each class $c$ the centroid:
\[
  \bar{\mathbf{h}}_c = \frac{1}{|\mathcal{I}_c|} \sum_{i \in \mathcal{I}_c} \tilde{\mathbf{h}}^{(i)},
\]
where $\mathcal{I}_c$ contains the indices of samples with label $c$. The variance loss is:
\[
  \mathcal{L}_\mathrm{var}
  = \frac{1}{C} \sum_{c=1}^C \frac{1}{|\mathcal{I}_c|} \sum_{i \in \mathcal{I}_c} 
    \left\| \tilde{\mathbf{h}}^{(i)} - \bar{\mathbf{h}}_c \right\|_2^2.
\]

\medskip
\noindent\textbf{Total Loss:}
The total objective combines the three components:
\[
  \mathcal{L}
  = \mathcal{L}_\mathrm{CE} + \mathcal{L}_\mathrm{contra} + \lambda\,\mathcal{L}_\mathrm{var},
\]
where we empirically set $\tau=0.07$ and $\lambda=0.1$ in our experiments.

\section{Datasets}
We examined our method on two brain MRI datasets: the ADNI dataset and an in-house dataset of headache patients. Both datasets consist of 3D brain MRIs. 

\subsection{Alzheimer’s Disease Neuroimaging Initiative Dataset}
The Alzheimer’s Disease Neuroimaging Initiative (ADNI) \href{http://adni.loni.usc.edu/}{(adni.loni.usc.edu)} is a longitudinal, multi-center observational study launched in 2004 to develop, standardize, and validate biomarkers for Alzheimer’s disease (AD) clinical trials. ADNI-1 initially enrolled 200 cognitively normal healthy controls (HC), 400 participants with mild cognitive impairment (MCI), and 200 with AD across 57 sites in the United States and Canada; later phases (ADNI-GO, ADNI-2 and ADNI-3) expanded the total cohort to over 1,000 subjects aged 55–90. Participants underwent serial neuroimaging (structural MRI T1-weighted MP-RAGE on 1.5T and 3T scanners; FDG-PET and amyloid PET tracers), biofluid collection (cerebrospinal fludi (CSF) and plasma biomarkers), genetic profiling (e.g., APOE genotyping), and comprehensive neuropsychological assessments (mini mental status exam (MMSE), Clinical Dementia Rating (CDR), Alzheimer’s Disease Assessment Scale – Cognitive Subscale (ADAS-Cog)) at baseline and follow-up visits (6, 12, 18, 24 months in ADNI-1; extended in subsequent phases). For our analyses, all T1-weighted MRI scans were preprocessed using non-linear registration to the MNI-152 template, N4 bias-field correction, skull-stripping, and intensity normalization via histogram matching. The dataset consists of 3 classes of images: healthy (HC), MCI, and AD. HC participants are participants with no subjective or objective memory complaints and normal performance on neuropsychological tests. Mild Cognitive Impairment (MCI) is the prodromal, intermediate stage between healthy aging and AD, characterized by subjective memory complaints, objective memory impairment (MMSE 24–30), CDR = 0.5, preserved activities of daily living, and absence of dementia. In our final processed dataset, we have 4769 T1-weighted brain MRI scans: 1831 HC, 1668 MCI, 1270 AD. 

\subsection{Institutional Headache Dataset}
Headache data were collected via prospective research approved by the  Institutional Review Board (IRB), and all participants provided written informed consent for their participation. At the time of enrollment, migraine participants were diagnosed with episodic or chronic migraine, with or without aura, based on the most recent edition of the International Classification of Headache Disorders (ICHD-3 beta or ICHD-3) \cite{olesen2008international}. Participants with acute post-traumatic headache (APTH) or persistent post-traumatic headache (PPTH) had PTH attributed to mild traumatic brain injury (mTBI) according to the latest ICHD criteria (ICHD-3 beta or ICHD-3). Individuals with a history of moderate or severe traumatic brain injury were excluded from the study. We collected MRIs of 96 individuals with migraine, 48 with APTH, 49 with PPTH, and 104 healthy controls from the institution. We extended our dataset by including MRIs of 428 healthy controls from the publicly available IXI dataset \cite{ixi}. For our experiments, we trained our model by first combining all headache types into one group and then investigated each subgroup’s performance separately. All 3D MRIs in this dataset were registered to the MNI152 1mm template and skull stripped. 
% \section{Results}
% We performed the experiments on two datasets. The first one is the publicly available ADNI (Cite) dataset that contain 3D MRI scan of 3 different cohorts of patients. The second one is our private dateset from the Mayo Clinic, which contains 3D MRI scans of different headache patients.  We performed the experiments to get the performance in binary classfication task. For the ADNI dataset we performed 3 settings of experiments. Normal vs AD, 

\section{Experiments and Results}
\label{sec:results}

We evaluated our slice-wise soft attention aggregation of DinoV2 features on two different tasks: Alzheimer’s disease detection using the ADNI dataset and headache detection on our private headache dataset. For the ADNI dataset, we performed three pairwise anomaly or disease detection tasks: HC vs.\ AD, HC vs.\ MCI, and MCI vs.\ AD.  
 For each task, we split the data with corresponding classes in an 80:10:10 ratio for training, validation, and testing.

 For the headache dataset, containing HC, migraine (MIG), APTH, and PPTH.  We performed the experiments in 2 different settings. Firstly, we performed experiments with HC vs. different headache types to evaluate the performance of headache detection using the brain MRIs. We trained four separate models to evaluate the method with four different scenarios: HC vs. all headache (considering all headache types: migraine, APTH, PPTH as one class), HC vs. Mig, HC vs. APTH, and HC vs. PPTH. Secondly, we perform experiments with a view to differentiating between the subtypes of headache: Mig. vs. APTH. Mig. vs PPTH and APTH vs PPTH. For HC vs. all headaches, we split the dataset with an 80:10:10 ratio for training, validation, and testing. For the HC vs. subtype scenarios, we put 10 samples from each class in the validation set to avoid bias during evaluation and performed 5-fold cross-validation, reporting the mean values. The results are shown in section~\ref{headache_results}.

In each case, we report binary classification performance in terms of accuracy, area under the ROC curve (AUC), along with an F$_1$ score, False Negative Rate (FNR), and we visualize confusion matrices. We also compare the results with two baseline methods: SC-MIL \cite{yan2023mil3d} and 3D ResNet to demonstrate the effectiveness of our method over them. 

% \begin{table}[H]
%   \centering
%   \caption{Binary classification performance on ADNI.}
%   \label{tab:adni_summary}
%   \resizebox{\columnwidth}{!}{
%   \begin{tabular}{llccc}
%     \toprule
%     Task        & Method & Accuracy (\%) & AUC & F$_1$ \\
%     \midrule
%     \multirow{3}{*}{HC vs.\ AD} 
                
%                 & SC-MIL      & 59.16 & 0.502 & 0.031 \\
%                 & ResNet 3D      & 84.35 & 0.854 & 0.816 \\
%                 & \textbf{DinoAtten3D} & \textbf{87.80} & \textbf{0.865} & \textbf{0.871} \\
%     \midrule
%     \multirow{3}{*}{HC vs.\ MCI} 
                
%                 & SC-MIL      & 48.29 & 0.476 & 0.385 \\
%                 & ResNet 3D       & 62.29 & 0.668 & 0.042 \\
%                 & \textbf{DinoAtten3D}  & \textbf{70.50} & \textbf{0.702} & \textbf{0.700} \\
%     \midrule
%     \multirow{3}{*}{MCI vs.\ AD} 
                
%                 & SC-MIL      & 56.46 & 0.514 & 0.218 \\
%                 & ResNet 3D       & 74.35 & 0.714 & 0.610 \\
%                 & \textbf{DinoAtten3D} & \textbf{75.70} & \textbf{0.749} & \textbf{0.730} \\
%     \bottomrule
%   \end{tabular}
%   }
% \end{table}

\begin{table}[H]
  \centering
  \caption{Binary classification performance on ADNI.}
  \label{tab:adni_summary}
  \resizebox{\columnwidth}{!}{
  \begin{tabular}{llcccc}
    \toprule
    Task        & Method & Accuracy (\%) & AUC & F$_1$ & FNR (\%) \\
    \midrule
    \multirow{3}{*}{HC vs.\ AD} 
                & SC-MIL      & 59.16 & 0.502 & 0.031 & 97.42 \\
                & ResNet 3D   & 84.35 & 0.854 & 0.816 & 23.62 \\
                & \textbf{DinoAtten3D} & \textbf{87.80} & \textbf{0.865} & \textbf{0.871} & \textbf{22.05} \\
    \midrule
    \multirow{3}{*}{HC vs.\ MCI} 
                & SC-MIL      & 48.29 & 0.476 & 0.385 & 65.87 \\
                & ResNet 3D   & 62.29 & 0.668 & 0.420 & 71.40 \\
                & \textbf{DinoAtten3D}  & \textbf{70.50} & \textbf{0.702} & \textbf{0.700} & \textbf{23.95} \\
    \midrule
    \multirow{3}{*}{MCI vs.\ AD} 
                & SC-MIL      & 56.46 & 0.514 & 0.218 & 85.75 \\
                & ResNet 3D   & 74.35 & 0.714 & 0.610 & 53.54 \\
                & \textbf{DinoAtten3D} & \textbf{75.70} & \textbf{0.749} & \textbf{0.730} & 
                \textbf{29.92}
                \\
    \bottomrule
  \end{tabular}
  }
\end{table}

\subsection{Alzheimer’s detection on ADNI dataset}

\begin{figure}[htbp]
  \centering
  % first panel (~32% of column width)
  \begin{subfigure}[b]{0.32\linewidth}
    \centering
    \includegraphics[width=\linewidth]{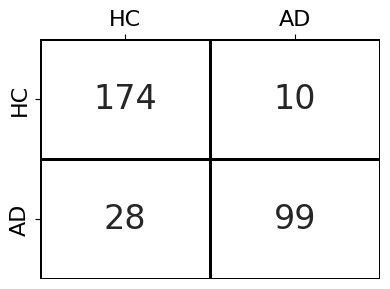}
    \caption{}
    \label{fig:cn_ad}
  \end{subfigure}\hfill
  % second panel
  \begin{subfigure}[b]{0.32\linewidth}
    \centering
    \includegraphics[width=\linewidth]{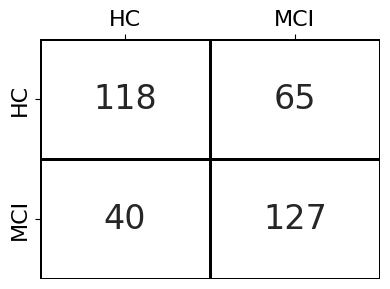}
    \caption{}
    \label{fig:cn_mci}
  \end{subfigure}\hfill
  % third panel
  \begin{subfigure}[b]{0.32\linewidth}
    \centering
    \includegraphics[width=\linewidth]{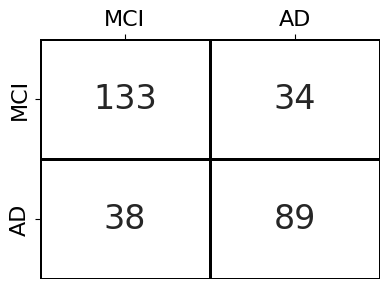}
    \caption{}
    \label{fig:mci_ad}
  \end{subfigure}
  \caption{Confusion matrix for ADNI dataset experiments: (a) HC vs AD, (b) HC vs MCI, (c) MCI vs AD}
  \label{fig:three_panels}
\end{figure}

Table~\ref{tab:adni_summary} summarizes results on three pairwise anomaly or disease detection tasks in the ADNI cohort: HC vs.\ AD, HC vs.\ MCI, and MCI vs.\ AD.  
Figures~\ref{fig:cn_ad}–\ref{fig:mci_ad} display the corresponding confusion matrices.

% \begin{table}[ht]
%   \centering
%   \caption{Binary classification performance on ADNI.}
%   \label{tab:adni_summary}
%   \resizebox{\columnwidth}{!}{
%   \begin{tabular}{lccc}
%     \toprule
%     Task        & Accuracy (\%) & AUC (\%) & Macro-F$_1$ (\%) \\
%     \midrule
%     HC vs.\ AD  & 87.80          & 86.50        & 87.10             \\
%     HC vs.\ MCI & 70.50          & 70.25     & 70.00             \\
%     MCI vs.\ AD & 75.70          & 74.90    & 73.00             \\
%     \bottomrule
%   \end{tabular}
%   }
  
% \end{table}

For HC vs.\ AD, our model achieves 87.80\% accuracy with an F$_1$ score of 0.871 in the test set after selecting the model with the lowest validation loss.  
As shown in Figure~\ref{fig:cn_ad}, 174/184 HC and 99/127 AD scans are correctly classified with 0.865 AUC, indicating good convergence and limited overfitting in terms of Alzheimer's disease detection. It also achieves better accuracy and AUC compared to the baselines. 

In the HC vs.\ MCI task, we obtain 70.50\% accuracy (AUC 0.702), with an F$_1$ score of 0.700.   
The confusion matrix in Figure~\ref{fig:cn_mci} shows moderate misclassifications (133 true HC vs.\ 38 false negatives; 89 true MCI vs.\ 34 false positives), reflecting the subtlety of early‐stage MCI detection. 

For MCI vs.\ AD, accuracy reaches 75.70\% (AUC 0.749), with an F$_1$ score of 0.730 for both classes.  
Figure~\ref{fig:mci_ad} illustrates that 133 MCI and 89 AD volumes are correctly labeled.

\begin{figure*}[ht]
  \centering
  % first panel (~32% of column width)
  \begin{subfigure}[b]{0.32\linewidth}
    \centering
    \includegraphics[width=\linewidth]{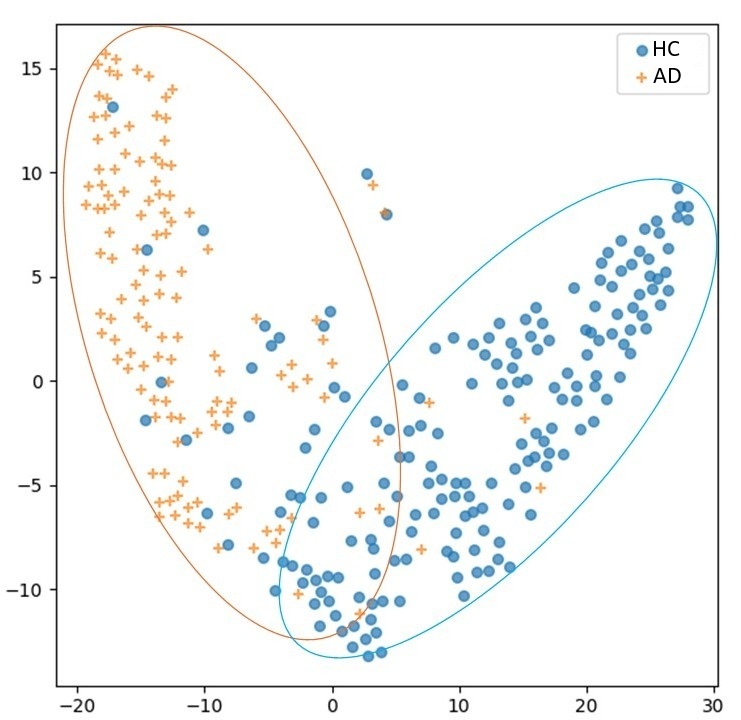}
    \caption{HC vs AD}
    \label{fig:tsne_adhc}
  \end{subfigure}\hfill
  % second panel
  \begin{subfigure}[b]{0.32\linewidth}
    \centering
    \includegraphics[width=\linewidth]{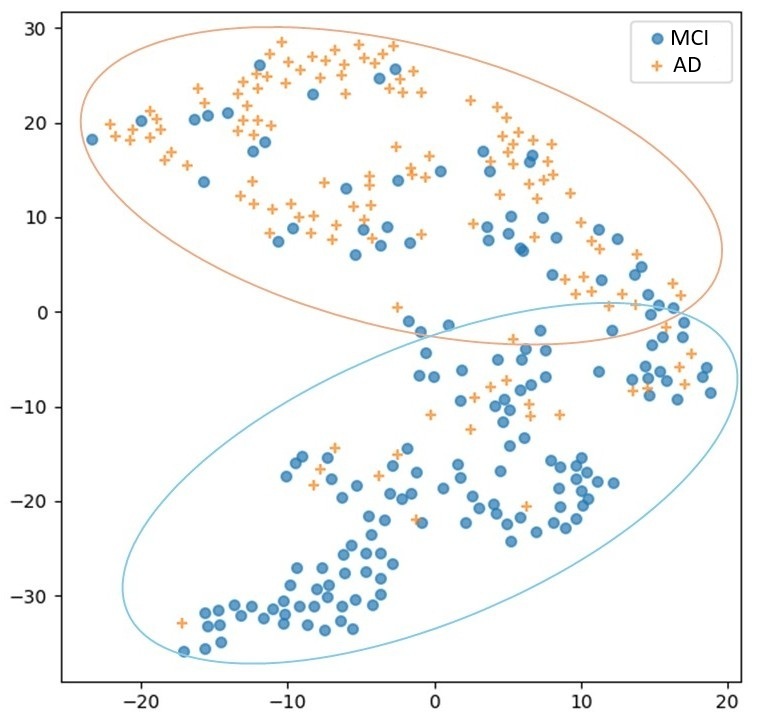}
    \caption{MCI vs AD}
    \label{fig:tsne_mciad}
  \end{subfigure}\hfill
  % third panel
  \begin{subfigure}[b]{0.32\linewidth}
    \centering
    \includegraphics[width=\linewidth]{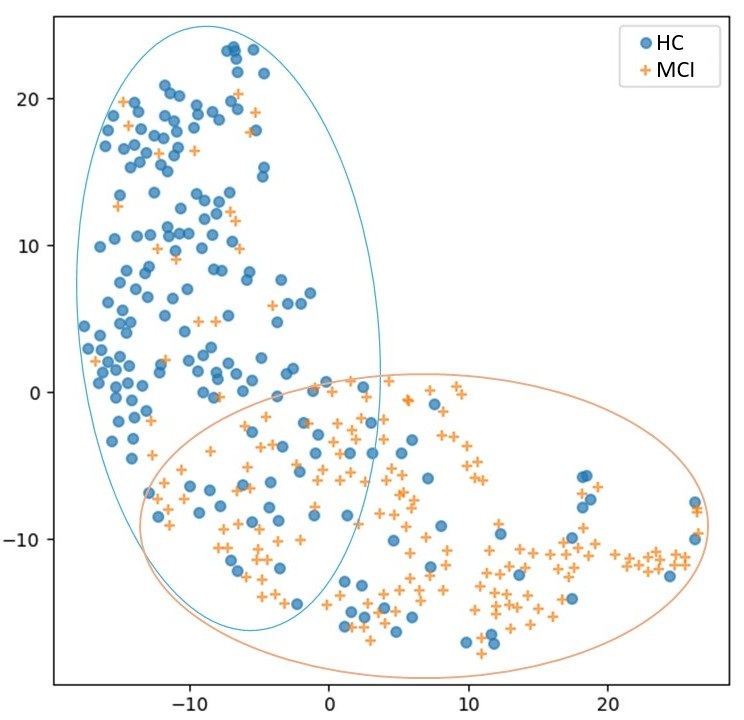}
    \caption{HC vs MCI}
    \label{fig:tsne_hcmci}
  \end{subfigure}
  \caption{t-SNE plots of Aggregated embeddings for ADNI dataset test subjects}
  \label{fig:three_panels}
\end{figure*}

As illustrated in Figures \ref{fig:tsne_adhc}–\ref{fig:tsne_hcmci}, the t-SNE projections of the aggregated patient embeddings for each ADNI classification task exhibit markedly different clustering behaviors. In the HC vs AD comparison (Figure \ref{fig:tsne_adhc}), the two cohorts form two well‐demarcated clusters, indicating a high degree of separability in the learned representation space. By contrast, the MCI vs AD (Figure \ref{fig:tsne_mciad}) and HC vs MCI (Figure \ref{fig:tsne_hcmci}) tasks display only moderate separation: while the embeddings tend to form cluster-like structures, there remains a non-negligible degree of overlap between classes. This suggests that, although the model captures discriminative features in all three scenarios, the boundary between mild cognitive impairment and either healthy controls or Alzheimer’s patients is less distinct than that between healthy and Alzheimer’s subjects.

\subsection{Headache classification on private dataset}
\label{headache_results}

\begin{table}[ht]
  \centering
  \caption{Binary classification performance on the headache dataset.}
  \label{tab:headache_summary}
  \resizebox{\columnwidth}{!}{ % fully utilizes text width
  \begin{tabular}{llcccc}
    \toprule
    Task & Method & Accuracy (\%) & AUC  & F$_1$ & FNR (\%) \\
    \midrule
    \multirow{3}{*}{HC vs.\ Headache}
        
         & SC-MIL           & 67.67 & 0.565 & 0.095 & 40.00\\
         & ResNet3D         & 82.18 & 80.96 & 0.705 & 35.00 \\
          & \textbf{DinoAtten3D} & \textbf{86.30} & \textbf{0.874} & \textbf{0.867} & \textbf{10.00}\\
    \midrule
    \multirow{3}{*}{HC vs.\ MIG}
         
         & SC-MIL           & 50.00 & 0.544 & 0.000 & 100.0\\
         & ResNet3D         & 74.05 & 0.870 & 0.696 & 40.00 \\
         & \textbf{DinoAtten3D} & \textbf{90.00} & \textbf{0.992} & \textbf{0.899} & \textbf{15.00}\\
    \midrule
    \multirow{3}{*}{HC vs.\ APTH}
         
         & SC-MIL           & 50.00  & 0.505  & 0.000 & 100.0 \\
         & ResNet3D         & 62.00  & 0.850  & 0.385 & 76.00 \\
         & \textbf{DinoAtten3D} &\textbf{ 85.00} & \textbf{0.970} & \textbf{0.846} & \textbf{30.00}\\
    \midrule
    \multirow{3}{*}{HC vs.\ PPTH}
         
         & SC-MIL           & 50.00 &  0.570 & 0.000  & 100.0\\
         & ResNet3D         & 65.00  & 0.828  & 0.565 & 54.10 \\
         & \textbf{DinoAtten3D} & \textbf{90.00} & \textbf{0.980} & \textbf{0.899} & \textbf{20.00}\\
    \bottomrule
  \end{tabular}
  }
\end{table}

\begin{figure*}[htbp]
  \centering
  \begin{minipage}{0.75\textwidth} % total width is 80% of text width
    \centering
    % four panels inside the 0.8\textwidth container
    \begin{subfigure}[b]{0.23\linewidth} % relative to minipage width
      \centering
      \includegraphics[width=\linewidth]{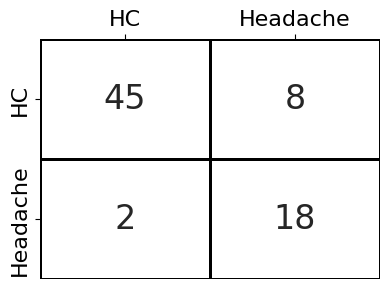}
      \caption{}
      \label{fig:hc_all}
    \end{subfigure}\hfill
    \begin{subfigure}[b]{0.23\linewidth}
      \centering
      \includegraphics[width=\linewidth]{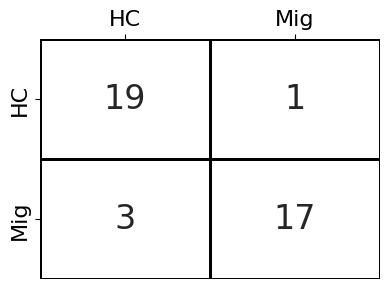}
      \caption{}
      \label{fig:hc_mig}
    \end{subfigure}\hfill
    \begin{subfigure}[b]{0.23\linewidth}
      \centering
      \includegraphics[width=\linewidth]{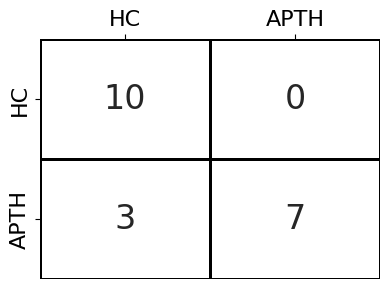}
      \caption{}
      \label{fig:hc_apth}
    \end{subfigure}\hfill
    \begin{subfigure}[b]{0.23\linewidth}
      \centering
      \includegraphics[width=\linewidth]{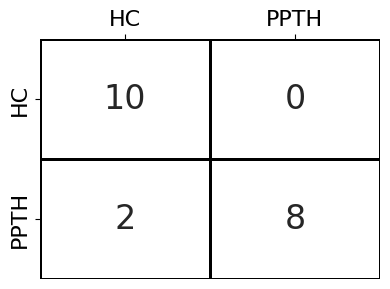}
      \caption{ }
      \label{fig:hc_ppth}
    \end{subfigure}
  \end{minipage}
  \caption{Confusion matrices for the four headache detection tasks. (a) HC vs All Headache, (b) HC vs Mig, (c) HC vs APTH, (d) HC vs PPTH}
  \label{fig:all_confusion}
\end{figure*}

We further tested our method on a private headache MRI dataset for headache detection and inter-headache classification. For the HC vs all headache scenario, the model achieved 86.30\% accuracy in the blind test set with 0.874 AUC. In the case of migraine detection, the model achieved 90.00\% accuracy with 0.993 AUC.  For APTH and PPTH detection with respect to HC, the model also achieved 85.00\% accuracy with 0.970 AUC and 90.00\% accuracy with 0.980 AUC, respectively. 
Table~\ref{tab:headache_summary} and Figures~\ref{fig:hc_all}–\ref{fig:hc_ppth} report the results and confusion matrix for headache detection.

% \begin{figure*}[htbp]
%   \centering
  
%   % four panels, each ~23% of text width, with equal spacing
%   \begin{subfigure}[b]{0.23\textwidth}
%     \centering
%     \includegraphics[width=\linewidth]{fig/HCvsHeadache.png}
%     \caption{HC vs all headache}
%     \label{fig:hc_all}
%   \end{subfigure}\hfill
%   \begin{subfigure}[b]{0.23\textwidth}
%     \centering
%     \includegraphics[width=\linewidth]{fig/HCvsMig.png}
%     \caption{HC vs Mig}
%     \label{fig:hc_mig}
%   \end{subfigure}\hfill
%   \begin{subfigure}[b]{0.23\textwidth}
%     \centering
%     \includegraphics[width=\linewidth]{fig/HCvsAPTH.png}
%     \caption{HC vs APTH}
%     \label{fig:hc_apth}
%   \end{subfigure}\hfill
%   \begin{subfigure}[b]{0.23\textwidth}
%     \centering
%     \includegraphics[width=\linewidth]{fig/HCvsPPTH.png}
%     \caption{HC vs PPTH}
%     \label{fig:hc_ppth}
%   \end{subfigure}
%   \caption{Confusion matrices for the four headache detection tasks.}
%   \label{fig:all_confusion}
% \end{figure*}

To gauge our model’s ability to detect anomalies across different headache subtypes, we further trained it on three binary tasks: Migraine (Mig) vs. Acute Post-Traumatic Headache (APTH), Persistent Post-Traumatic Headache (PPTH) vs. APTH, and Mig vs. PPTH. Table \ref{tab:intr_headache_summary} reports, for each task, the validation accuracy, F$_1$-score, and AUC. For Mig. vs APTH and APTH vs PPTH, the model achieved high accuracy of 90\% and 95\% with almost perfect AUC scores. However, for the Mig vs PPTH classification, the model achieved 55\% accuracy, close to the best-performing baseline method in terms of accuracy and AUC.

 \begin{table}[ht]
  \centering
  \caption{Performance of DinoAtten for inter-headache classification.}
  \label{tab:intr_headache_summary}
  \resizebox{\columnwidth}{!}{
   \begin{tabular}{llccc}
    \toprule
    Task & Method & Accuracy (\%) & AUC  & F$_1$  \\
    \midrule
    \multirow{3}{*}{MIG vs.\ APTH}
         
         & SC-MIL           & 50.00 &  0.388 & 0.667   \\
         & ResNet3D         & 78.01    & 0.920    & 0.798    \\
         & \textbf{DinoAtten3D} & \textbf{90.00} & \textbf{0.980} & \textbf{0.890} \\
    \midrule
    \multirow{3}{*}{APTH vs.\ PPTH}
         
         & SC-MIL           & 50.00 &  0.540 & 0.667   \\
         & ResNet3D         & 84.00    & 0.989    & 0.802    \\
         & \textbf{DinoAtten3D} & \textbf{95.00} & \textbf{0.991} & \textbf{0.949} \\
    \midrule
    \multirow{3}{*}{MIG vs.\ PPTH}
         
         & SC-MIL            & 50.00 &  0.520 & 0.000  \\
         & ResNet3D         & \textbf{58.00}    & 0.562    & \textbf{0.698}    \\
         & \textbf{DinoAtten3D} & 55.00 & 0.680 & 0.436 \\
    \bottomrule
  \end{tabular}
  }
\end{table}

      % put this in your preamble
                  % no extra space above sub-captions

\begin{figure*}[htbp]
  \centering
  % first row, two panels at ~48% of column width
  \begin{subfigure}[b]{0.48\linewidth}
    \centering
    \includegraphics[width=\linewidth]{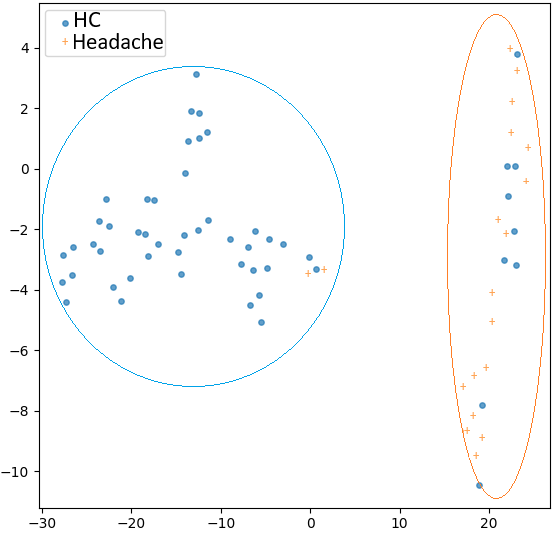}
    \caption{HC vs All Headache}
    \label{fig:tsne_hc_hdc}
  \end{subfigure}\hfill
  \begin{subfigure}[b]{0.48\linewidth}
    \centering
    \includegraphics[width=\linewidth]{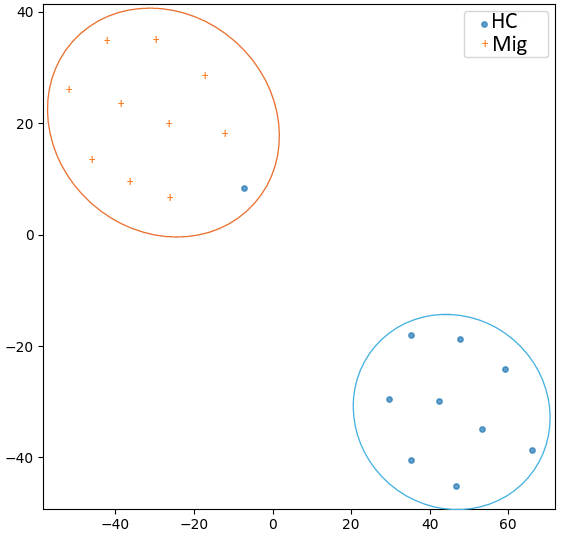}
    \caption{HC vs Mig}
    \label{fig:tsne_hcmig}
  \end{subfigure}\\[1em]
  % second row, two more panels
  \begin{subfigure}[b]{0.48\linewidth}
    \centering
    \includegraphics[width=\linewidth]{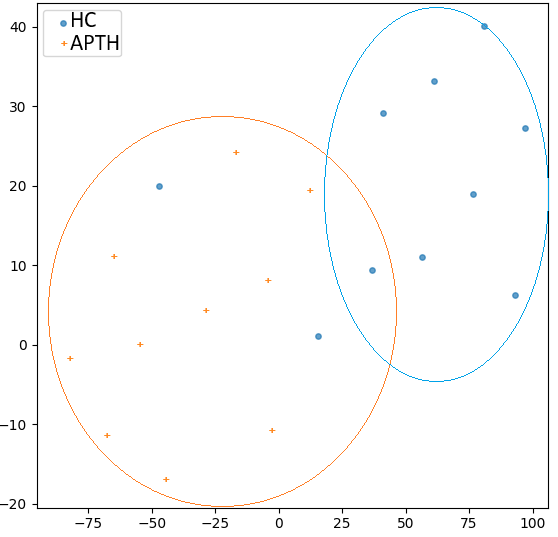}
    \caption{HC vs APTH}
    \label{fig:tsne_hcapth}
  \end{subfigure}\hfill
  \begin{subfigure}[b]{0.48\linewidth}
    \centering
    \includegraphics[width=\linewidth]{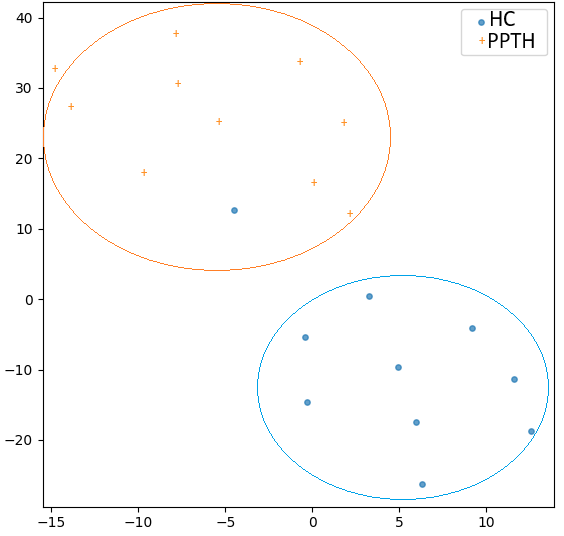}
    \caption{HC vs PPTH}
    \label{fig:tsne_hcppth}
  \end{subfigure}
  \caption{t-SNE plots of aggregated embeddings for Headache dataset test subjects.}
  \label{fig:four_panels}
\end{figure*}

Figures \ref{fig:tsne_hc_hdc}–\ref{fig:tsne_hcppth} depict the t-SNE projections of the aggregated embeddings for test subjects in the headache dataset experiments. The results reveal well-defined clusters corresponding to healthy controls versus the various headache subtypes. In particular, the HC versus Migraine, HC versus PPTH, and HC versus all-headache comparisons exhibit pronounced separation in embedding space, with less inter-cluster overlap. While the HC versus APTH plot shows a modest degree of intermingling, the two groups remain largely distinguishable. These observations corroborate the efficacy of incorporating supervised contrastive learning alongside class-variance regularization during model training, as they yield representations that enhance class discriminability across headache detection tasks.
\section{Discussion}
Across the binary tasks, our attention-weighted DinoV2 embeddings exhibit consistently strong anomaly detection when the classes are clearly distinct (i.e., AD vs. HC, all headache types vs HC, Mig vs APTH, APTH vs PPTH), but struggle when they share similar features (e.g., PPTH vs Mig). In the Mig vs APTH experiment, the model achieved 90\% accuracy, a 0.89 F$_1$-score, and a 0.98 AUC, demonstrating robust separation of acute post-traumatic headache as the anomalous class. Performance improved slightly in the PPTH vs APTH comparison (95\% accuracy, 0.95 F$_1$-score, 0.99 AUC), indicating almost perfect discrimination of persistent post-traumatic headache against acute cases. By contrast, the Migraine vs PPTH task yielded only 55\% accuracy, a 0.44 F$_1$-score, and a 0.68 AUC, reflecting the model’s difficulty in distinguishing post-traumatic profiles with Migraine. This finding reflects the clinical observation that symptoms of Mig and PPTH are typically very similar, although the symptoms of PPTH are triggered by a brain injury, whereas those of migraine are not. These results demonstrate that our aggregation strategy excels at detecting anomalies when the target condition is well-defined, yet also highlight the need for further feature refinement in cases of subtly differing health conditions. In the future, we plan to comprehensively assess our approach in more challenging scenarios where data scarcity and class imbalance are more prevalent. 

\section{Conclusion}
In this study, we present a slice-level attention aggregation framework built upon DinoV2, a self-supervised vision transformer pretrained predominantly on natural images. Despite the domain shift, our method demonstrates strong performance across both neuroimaging and headache classification tasks. Notably, it achieves high accuracy in distinguishing healthy from pathological cases and shows competitive performance even in challenging inter-subtype classifications. The results on the ADNI dataset underscore the model’s capacity to reliably detect Alzheimer's disease, while the promising outcomes in differentiating among headache subtypes (e.g., migraine vs.\ post-traumatic headache) highlight the discriminative power of the learned representations. Our findings suggest that the rich, generalized features extracted by DinoV2 are transferable to medical imaging contexts, enabling robust anomaly classification even in data-scarce scenarios. The effectiveness of our slice-level soft attention mechanism further validates the importance of localized features in volumetric medical data. Future work will explore leveraging these features for more granular multi-class disease sub-typing and investigating domain-adaptive pretraining strategies to further enhance performance in specialized clinical applications.

% \textbf{Acknowledgment: } This work was supported by the United States Department of Defense W81XWH-15-1-0286 and W81XWH1910534, National Institutes of Health K23NS070891, National Institutes of Health—National Institute of Neurological Disorders and Stroke, Award Number 1R61NS113315-01, and Amgen Investigator Sponsored Study 20187183.
{
    \small
    \bibliographystyle{ieeenat_fullname}
    \bibliography{main}

\begin{thebibliography}{43}
\providecommand{\natexlab}[1]{#1}
\providecommand{\url}[1]{\texttt{#1}}
\expandafter\ifx\csname urlstyle\endcsname\relax
  \providecommand{\doi}[1]{doi: #1}\else
  \providecommand{\doi}{doi: \begingroup \urlstyle{rm}\Url}\fi

\bibitem[Baharoon et~al.(2023)Baharoon, Qureshi, Ouyang, Xu, Aljouie, and Peng]{baharoon2023evaluating}
Mohammed Baharoon, Waseem Qureshi, Jiahong Ouyang, Yanwu Xu, Abdulrhman Aljouie, and Wei Peng.
\newblock Evaluating general purpose vision foundation models for medical image analysis: An experimental study of dinov2 on radiology benchmarks.
\newblock \emph{arXiv preprint arXiv:2312.02366}, 2023.

\bibitem[Bommasani et~al.(2021)Bommasani, Hudson, Adeli, Altman, Arora, von Arx, Bernstein, Bohg, Bosselut, Brunskill, et~al.]{bommasani2021opportunities}
Rishi Bommasani, Drew~A Hudson, Ehsan Adeli, Russ Altman, Simran Arora, Sydney von Arx, Michael~S Bernstein, Jeannette Bohg, Antoine Bosselut, Emma Brunskill, et~al.
\newblock On the opportunities and risks of foundation models.
\newblock \emph{arXiv preprint arXiv:2108.07258}, 2021.

\bibitem[Brain Development()]{ixi}
Brain Development.
\newblock Ixi dataset.
\newblock \url{https://brain-development.org/ixi-dataset/}.

\bibitem[Campanella et~al.(2019)Campanella, Hanna, Geneslaw, Miraflor, Werneck Krauss~Silva, Busam, Brogi, Reuter, Klimstra, and Fuchs]{campanella2019clinical}
Gabriele Campanella, Matthew~G Hanna, Luke Geneslaw, Allen Miraflor, Vitor Werneck Krauss~Silva, Klaus~J Busam, Edi Brogi, Victor~E Reuter, David~S Klimstra, and Thomas~J Fuchs.
\newblock Clinical-grade computational pathology using weakly supervised deep learning on whole slide images.
\newblock \emph{Nature medicine}, 25\penalty0 (8):\penalty0 1301--1309, 2019.

\bibitem[Carbonneau et~al.(2018)Carbonneau, Cheplygina, Granger, and Gagnon]{carbonneau2018multiple}
Marc-Andre Carbonneau, Veronika Cheplygina, Eric Granger, and Guillaume Gagnon.
\newblock Multiple instance learning: A survey of problem characteristics and applications.
\newblock \emph{Pattern Recognition}, 77:\penalty0 329--353, 2018.

\bibitem[Che et~al.(2025)Che, Rafsani, Shah, Siddiquee, and Wu]{che2025anofpdm}
Yiming Che, Fazle Rafsani, Jay Shah, Md~Mahfuzur~Rahman Siddiquee, and Teresa Wu.
\newblock Anofpdm: Anomaly detection with forward process of diffusion models for brain mri.
\newblock In \emph{Proceedings of the Winter Conference on Applications of Computer Vision}, pages 1113--1122, 2025.

\bibitem[Ciga et~al.(2021)Ciga, Xu, Nofech-Mozes, Noy, Lu, and Martel]{ciga2021overcoming}
Ozan Ciga, Tony Xu, Sharon Nofech-Mozes, Shawna Noy, Fang-I Lu, and Anne~L Martel.
\newblock Overcoming the limitations of patch-based learning to detect cancer in whole slide images.
\newblock \emph{Scientific Reports}, 11\penalty0 (1):\penalty0 8894, 2021.

\bibitem[Crespi et~al.(2022)Crespi, Loiacono, and Sartori]{crespi20223d}
Leonardo Crespi, Daniele Loiacono, and Pierandrea Sartori.
\newblock Are 3d better than 2d convolutional neural networks for medical imaging semantic segmentation?
\newblock In \emph{2022 International Joint Conference on Neural Networks (IJCNN)}, pages 1--8. IEEE, 2022.

\bibitem[Daras et~al.(2024)Daras, Tang, Tang, et~al.]{daras2024how}
Giannis Daras, Yuyang Tang, Shiyu Tang, et~al.
\newblock How much is a noisy image worth? data scaling laws for ambient diffusion.
\newblock \emph{arXiv preprint arXiv:2401.03196}, 2024.

\bibitem[Deepak and Ameer(2019)]{deepak2019brain}
S Deepak and PM Ameer.
\newblock Brain tumor classification using deep cnn features via transfer learning.
\newblock \emph{Computers in biology and medicine}, 111:\penalty0 103345, 2019.

\bibitem[Dhariwal and Nichol(2021)]{dhariwal2021diffusion}
Prafulla Dhariwal and Alexander Nichol.
\newblock Diffusion models beat gans on image synthesis.
\newblock \emph{Advances in neural information processing systems}, 34:\penalty0 8780--8794, 2021.

\bibitem[Havaei et~al.(2017)Havaei, Davy, Warde-Farley, Biard, Courville, Bengio, Pal, Jodoin, and Larochelle]{havaei2017brain}
Mohammad Havaei, Axel Davy, David Warde-Farley, Antoine Biard, Aaron Courville, Yoshua Bengio, Chris Pal, Pierre-Marc Jodoin, and Hugo Larochelle.
\newblock Brain tumor segmentation with deep neural networks.
\newblock \emph{Medical image analysis}, 35:\penalty0 18--31, 2017.

\bibitem[Huix et~al.(2024)Huix, Ganeshan, Haslum, S{\"o}derberg, Matsoukas, and Smith]{huix2024natural}
Joana~Pal{\'e}s Huix, Adithya~Raju Ganeshan, Johan~Fredin Haslum, Magnus S{\"o}derberg, Christos Matsoukas, and Kevin Smith.
\newblock Are natural domain foundation models useful for medical image classification?
\newblock In \emph{Proceedings of the IEEE/CVF winter conference on applications of computer vision}, pages 7634--7643, 2024.

\bibitem[Ilse et~al.(2018)Ilse, Tomczak, and Welling]{ilse2018attention}
Max Ilse, Jakub~M Tomczak, and Max Welling.
\newblock Attention-based deep multiple instance learning.
\newblock In \emph{International conference on machine learning}, pages 2127--2136. PMLR, 2018.

\bibitem[Iqbal et~al.(2018)Iqbal, Ghani, Saba, and Rehman]{iqbal2018brain}
Sajid Iqbal, M~Usman Ghani, Tanzila Saba, and Amjad Rehman.
\newblock Brain tumor segmentation in multi-spectral mri using convolutional neural networks (cnn).
\newblock \emph{Microscopy research and technique}, 81\penalty0 (4):\penalty0 419--427, 2018.

\bibitem[Kayalibay et~al.(2017)Kayalibay, Jensen, and van~der Smagt]{kayalibay2017cnn}
Baris Kayalibay, Grady Jensen, and Patrick van~der Smagt.
\newblock Cnn-based segmentation of medical imaging data.
\newblock \emph{arXiv preprint arXiv:1701.03056}, 2017.

\bibitem[Li et~al.(2024)Li, Yu, Zhang, et~al.]{li2024pruning}
Haonan Li, Tete Yu, Xinyang Zhang, et~al.
\newblock Pruning then reweighting: Towards data-efficient training of diffusion models.
\newblock \emph{arXiv preprint arXiv:2401.04224}, 2024.

\bibitem[Li et~al.(2014)Li, Cai, Wang, Zhou, Feng, and Chen]{li2014medical}
Qing Li, Weidong Cai, Xiaogang Wang, Yun Zhou, David~Dagan Feng, and Mei Chen.
\newblock Medical image classification with convolutional neural network.
\newblock In \emph{2014 13th international conference on control automation robotics \& vision (ICARCV)}, pages 844--848. IEEE, 2014.

\bibitem[Li et~al.(2021)Li, Xu, Wang, Wang, Ma, and Zhang]{li2021scmil}
Xiang Li, Ming Xu, Bing Wang, Licheng Wang, Kaicheng Ma, and Shaoting Zhang.
\newblock Sc-mil: Supervised contrastive multiple instance learning for imbalanced classification in pathology.
\newblock In \emph{International Conference on Computer Vision (ICCV)}, pages 14971--14981, 2021.

\bibitem[Liu et~al.(2024)Liu, Zhang, Dai, Zhang, Cai, Zhou, and Chen]{liu2024few}
Fan Liu, Tianshu Zhang, Wenwen Dai, Chuanyi Zhang, Wenwen Cai, Xiaocong Zhou, and Delong Chen.
\newblock Few-shot adaptation of multi-modal foundation models: A survey.
\newblock \emph{Artificial Intelligence Review}, 57\penalty0 (10):\penalty0 268, 2024.

\bibitem[Lu et~al.(2021)Lu, Williamson, Chen, Chen, Barbieri, and Mahmood]{lu2021data}
Ming~Y Lu, Drew F~K Williamson, Tiffany~Y Chen, Ronald~J Chen, Milena Barbieri, and Faisal Mahmood.
\newblock Data-efficient and weakly supervised computational pathology on whole-slide images.
\newblock \emph{Nature Biomedical Engineering}, 5\penalty0 (6):\penalty0 555--570, 2021.

\bibitem[Ma et~al.(2024)Ma, He, Li, Han, You, and Wang]{ma2024segment}
Jun Ma, Yuting He, Feifei Li, Lin Han, Chenyu You, and Bo Wang.
\newblock Segment anything in medical images.
\newblock \emph{Nature Communications}, 15\penalty0 (1):\penalty0 654, 2024.

\bibitem[Mortazi and Bagci(2018)]{mortazi2018automatically}
Aliasghar Mortazi and Ulas Bagci.
\newblock Automatically designing cnn architectures for medical image segmentation.
\newblock In \emph{Machine Learning in Medical Imaging: 9th International Workshop, MLMI 2018, Held in Conjunction with MICCAI 2018, Granada, Spain, September 16, 2018, Proceedings 9}, pages 98--106. Springer, 2018.

\bibitem[Mykula et~al.(2024)Mykula, Gasser, Lobmaier, Schnabel, Zimmer, and Bercea]{mykula2024diffusion}
Hanna Mykula, Lisa Gasser, Silvia Lobmaier, Julia~A Schnabel, Veronika Zimmer, and Cosmin~I Bercea.
\newblock Diffusion models for unsupervised anomaly detection in fetal brain ultrasound.
\newblock In \emph{International Workshop on Advances in Simplifying Medical Ultrasound}, pages 220--230. Springer, 2024.

\bibitem[Olesen(2008)]{olesen2008international}
Jes Olesen.
\newblock The international classification of headache disorders.
\newblock \emph{Headache: The Journal of Head and Face Pain}, 48\penalty0 (5):\penalty0 691--693, 2008.

\bibitem[Oquab et~al.(2023)Oquab, Darcet, Moutakanni, Vo, Szafraniec, Khalidov, Fernandez, Haziza, Massa, El-Nouby, et~al.]{oquab2023dinov2}
Maxime Oquab, Timoth{\'e}e Darcet, Th{\'e}o Moutakanni, Huy Vo, Marc Szafraniec, Vasil Khalidov, Pierre Fernandez, Daniel Haziza, Francisco Massa, Alaaeldin El-Nouby, et~al.
\newblock Dinov2: Learning robust visual features without supervision.
\newblock \emph{arXiv preprint arXiv:2304.07193}, 2023.

\bibitem[Paul et~al.(2024)Paul, Islam, Rafsani, Khorasani, and Soumma]{paul2024efficient}
Plabon Paul, Md~Nazmul Islam, Fazle Rafsani, Pegah Khorasani, and Shovito~Barua Soumma.
\newblock Efficient feature extraction and classification architecture for mri-based brain tumor detection.
\newblock \emph{arXiv preprint arXiv:2410.22619}, 2024.

\bibitem[Radford et~al.(2018)Radford, Narasimhan, Salimans, Sutskever, et~al.]{radford2018improving}
Alec Radford, Karthik Narasimhan, Tim Salimans, Ilya Sutskever, et~al.
\newblock Improving language understanding by generative pre-training.
\newblock 2018.

\bibitem[Radford et~al.(2021)Radford, Kim, Hallacy, Ramesh, Goh, Agarwal, Sastry, Askell, Mishkin, Clark, et~al.]{radford2021learning}
Alec Radford, Jong~Wook Kim, Chris Hallacy, Aditya Ramesh, Gabriel Goh, Sandhini Agarwal, Girish Sastry, Amanda Askell, Pamela Mishkin, Jack Clark, et~al.
\newblock Learning transferable visual models from natural language supervision.
\newblock In \emph{International conference on machine learning}, pages 8748--8763. PmLR, 2021.

\bibitem[Rahman~Siddiquee et~al.(2022)Rahman~Siddiquee, Shah, Wu, Chong, Schwedt, and Li]{rahman2022healthygan}
Md~Mahfuzur Rahman~Siddiquee, Jay Shah, Teresa Wu, Catherine Chong, Todd Schwedt, and Baoxin Li.
\newblock Healthygan: Learning from unannotated medical images to detect anomalies associated with human disease.
\newblock In \emph{International Workshop on Simulation and Synthesis in Medical Imaging}, pages 43--54. Springer, 2022.

\bibitem[Salehi et~al.(2023)Salehi, Khan, Gupta, Alabduallah, Almjally, Alsolai, Siddiqui, and Mellit]{salehi2023study}
Ahmad~Waleed Salehi, Shakir Khan, Gaurav Gupta, Bayan~Ibrahimm Alabduallah, Abrar Almjally, Hadeel Alsolai, Tamanna Siddiqui, and Adel Mellit.
\newblock A study of cnn and transfer learning in medical imaging: Advantages, challenges, future scope.
\newblock \emph{Sustainability}, 15\penalty0 (7):\penalty0 5930, 2023.

\bibitem[Schlegl et~al.(2017)Schlegl, Seebock, Waldstein, Langs, and Schmidt-Erfurthb]{schlegl2017fast}
Thomas Schlegl, Philipp Seebock, Sebastian~M Waldstein, Georg Langs, and Ursula Schmidt-Erfurthb.
\newblock Fast unsupervised anomaly detection with generative adversarial networks.
\newblock \emph{Medical Image Analysis}, 2:\penalty0 2, 2017.

\bibitem[Shin et~al.(2016)Shin, Roth, Gao, Lu, Xu, Nogues, Yao, Mollura, and Summers]{shin2016deep}
Hoo-Chang Shin, Holger~R Roth, Mingchen Gao, Le Lu, Ziyue Xu, Isabella Nogues, Jianhua Yao, Daniel Mollura, and Ronald~M Summers.
\newblock Deep convolutional neural networks for computer-aided detection: Cnn architectures, dataset characteristics and transfer learning.
\newblock \emph{IEEE transactions on medical imaging}, 35\penalty0 (5):\penalty0 1285--1298, 2016.

\bibitem[Siddiquee et~al.(2024)Siddiquee, Shah, Wu, Chong, Schwedt, Dumkrieger, Nikolova, and Li]{siddiquee2024brainomaly}
Md~Mahfuzur~Rahman Siddiquee, Jay Shah, Teresa Wu, Catherine Chong, Todd~J Schwedt, Gina Dumkrieger, Simona Nikolova, and Baoxin Li.
\newblock Brainomaly: Unsupervised neurologic disease detection utilizing unannotated t1-weighted brain mr images.
\newblock In \emph{Proceedings of the IEEE/CVF Winter Conference on Applications of Computer Vision}, pages 7573--7582, 2024.

\bibitem[Tschuchnig and Gadermayr(2021)]{tschuchnig2021anomaly}
Maximilian~E Tschuchnig and Michael Gadermayr.
\newblock Anomaly detection in medical imaging-a mini review.
\newblock In \emph{International Data Science Conference}, pages 33--38. Springer, 2021.

\bibitem[Varoquaux and Cheplygina(2022)]{varoquaux2022machine}
Ga{\"e}l Varoquaux and Veronika Cheplygina.
\newblock Machine learning for medical imaging: methodological failures and recommendations for the future.
\newblock \emph{NPJ digital medicine}, 5\penalty0 (1):\penalty0 48, 2022.

\bibitem[Wang et~al.(2025)Wang, Alabdulmohsin, Salz, Li, Rong, and Zhai]{wang2025scaling}
Xiao Wang, Ibrahim Alabdulmohsin, Daniel Salz, Zhe Li, Keran Rong, and Xiaohua Zhai.
\newblock Scaling pre-training to one hundred billion data for vision language models.
\newblock \emph{arXiv preprint arXiv:2502.07617}, 2025.

\bibitem[Wang et~al.(2023)Wang, Jiang, Zheng, Wang, He, Wang, Chen, Zhou, et~al.]{wang2023patch}
Zhendong Wang, Yifan Jiang, Huangjie Zheng, Peihao Wang, Pengcheng He, Zhangyang Wang, Weizhu Chen, Mingyuan Zhou, et~al.
\newblock Patch diffusion: Faster and more data-efficient training of diffusion models.
\newblock \emph{Advances in neural information processing systems}, 36:\penalty0 72137--72154, 2023.

\bibitem[Wyatt et~al.(2022)Wyatt, Leach, Schmon, and Willcocks]{wyatt2022anoddpm}
Julian Wyatt, Adam Leach, Sebastian~M Schmon, and Chris~G Willcocks.
\newblock Anoddpm: Anomaly detection with denoising diffusion probabilistic models using simplex noise.
\newblock In \emph{Proceedings of the IEEE/CVF conference on computer vision and pattern recognition}, pages 650--656, 2022.

\bibitem[Yan et~al.(2023)Yan, Song, Xu, Wang, Wang, Ma, and Zhang]{yan2023mil3d}
Weilin Yan, Haoran Song, Ming Xu, Bing Wang, Licheng Wang, Kaicheng Ma, and Shaoting Zhang.
\newblock Mil3d: Multi-instance learning for 3d medical image classification.
\newblock In \emph{Medical Image Computing and Computer-Assisted Intervention (MICCAI)}, 2023.

\bibitem[Zhang et~al.(2022)Zhang, Guo, Zhang, Lu, and Zhao]{zhang2022unsupervised}
Haibo Zhang, Wenping Guo, Shiqing Zhang, Hongsheng Lu, and Xiaoming Zhao.
\newblock Unsupervised deep anomaly detection for medical images using an improved adversarial autoencoder.
\newblock \emph{Journal of Digital Imaging}, 35\penalty0 (2):\penalty0 153--161, 2022.

\bibitem[Zhang and Metaxas(2024)]{zhang2024challenges}
Shaoting Zhang and Dimitris Metaxas.
\newblock On the challenges and perspectives of foundation models for medical image analysis.
\newblock \emph{Medical image analysis}, 91:\penalty0 102996, 2024.

\bibitem[Zhang et~al.(2023)Zhang, Xu, Usuyama, Xu, Bagga, Tinn, Preston, Rao, Wei, Valluri, et~al.]{zhang2023biomedclip}
Sheng Zhang, Yanbo Xu, Naoto Usuyama, Hanwen Xu, Jaspreet Bagga, Robert Tinn, Sam Preston, Rajesh Rao, Mu Wei, Naveen Valluri, et~al.
\newblock Biomedclip: a multimodal biomedical foundation model pretrained from fifteen million scientific image-text pairs.
\newblock \emph{arXiv preprint arXiv:2303.00915}, 2023.

\end{thebibliography}
}

\end{document}